\documentclass[a4paper,12pt]{article}
\usepackage[cp1251]{inputenc}
\usepackage[english]{babel}
\inputencoding{cp1251}
\def\l{\lambda}
\def\m{\mu}

\def\d{\delta}

\def\a{\alpha}
\def\b{\beta}

\def\bba{\begin{array}}
\def\eea{\end{array}}

\def\bb{\begin{equation}}
\def\ee{\end{equation}}

\begin{document}

\author{Kazakova T.G.}

\title{Finite-dimensional reductions of the discrete Toda chain}

\date{\null}
\maketitle

\begin{abstract}
The problem of construction of integrable boundary conditions for
the discrete Toda chain is considered. The restricted chains for
properly chosen closure conditions are reduced to the well known
discrete Painlev\'e equations $dP_{III}$, $dP_{V}$, $dP_{VI}$.
Lax representations for these discrete Painlev\'e equations are
found.
\end{abstract}

\section{Introduction}

It is well known that the Toda lattice equation \bb\label{ut}
q_{n,xx}=e^{q_{n+1}-q_{n}}-e^{q_{n}-q_{n-1}} \ee admits several
different integrable discretizations \cite{hi}-\cite{su1}. Let us
consider one of them \cite{su}
$$
q_{m+1,n}-2q_{m,n}+q_{m-1,n}=\ln\frac{e^{q_{m,n+1}-q_{m,n}}+1}{e^{q_{m,n}-q_{m,n-1}}+1}
$$
which can be also presented in variables $u_{m,n}=e^{q_{m,n}}$
\bb\label{toda} u_{m+1,n}={u^2_{m,n}(1+u_{m,n+1}/u_{m,n})\over
u_{m-1,n}(1+u_{m,n}/u_{m,n-1})}. \ee The discrete Toda chain
(\ref{toda}) is referred also as 2-dimensional reduction of
Hirota's bilinear equation \cite{ward}, \cite{zab}, which has
applications in statistical physics and quantum field theory
\cite{crich}, \cite{kor}.

One of the most effective methods for constructing solutions of a
discrete chain is to find its integrable finite-dimensional
reductions. In most cases, this pertains to its periodic closure.
But there are other possibilities for truncating the chains while
conserving the integrability property \cite{sk}, \cite{su},
\cite{ah}. For chains that admit zero curvature representation,
there is a simple and effective method for seeking cut-off
constraints (boundary conditions) compatible with the conservation
laws of the chain \cite{hk}-\cite{kaz}.

The  discrete  Toda chain (\ref{toda}) is equivalent to the matrix
equation \bb\label{la}
L_{m+1,n}(\lambda)A_{m,n}(\lambda)=A_{m,n+1}(\lambda)L_{m,n}(\lambda),
\ee which is a consistency condition (the zero curvature equation)
of two linear equations
\begin{eqnarray}
&& Y_{m,n+1}(\lambda)=L_{m,n}(\lambda)Y_{m,n}(\lambda),\label{l}\\
&& Y_{m+1,n}(\lambda)=A_{m,n}(\lambda)Y_{m,n}(\lambda),\label{a}
\end{eqnarray}
where $\l$ is a parameter and $L_{m,n}$, $A_{m,n}$ are matrices of
the following form \cite{su}
$$
L_{m,n}=\left(\begin{array}{cc}
\lambda +{u_{m,n}\over u_{m-1,n}} & u_{m,n} \\
 \lambda {1\over u_{m-1,n}} & 0 \end {array}\right),\quad
A_{m,n}=\left(\begin{array}{cc} \lambda  & u_{m,n} \\ \lambda
{1\over u_{m,n-1}} & -1
 \end {array}\right).
$$

{\bf Definition.} {\it We will call a boundary condition
\bb\label{bc}
  u_{m,0}=F(m,u_{m,1},u_{m-1,1},...,u_{m,M},u_{m-1,M})
\ee compatible with zero curvature equation (\ref{la}) if equation
(\ref{a}) at the spatial point $n=1$ \bb\label{abc}
  Y_{m+1,1}(\lambda)=A_{m,1}(\lambda)\vert_{u_{m,0}=F}Y_{m,1}(\lambda)
\ee has an additional point symmetry of the form \bb\label{y_new}
  \tilde Y_{m,1}(\tilde \lambda)=H(m,[u],\lambda)Y_{m,1}(\lambda),
\quad \tilde \lambda=h(\lambda). \ee}

In other words boundary condition (\ref{bc}) is integrable if
there exists a matrix-valued function $$
H(m,[u],\lambda)=H(m,u_{m,1},u_{m-1,1},...,u_{m,k},u_{m-1,k},\lambda)
$$
together with the involution $\tilde \lambda=h(\lambda)$ such that
for any solution $Y_{m,0}(\lambda)$ of the equation (\ref{abc})
the function (\ref{y_new}) is a solution of  the same equation.
This means that the following equality \bb\label{hl}
H(m+1,[u],\lambda)A_{m,0}(\lambda)=A_{m,0}(\tilde
\lambda)H(m,[u],\lambda) \ee is valid.

We note that equation (\ref{hl}) contains three unknowns (the
boundary condition $F(m,[u])$, the involution $ \tilde{\l}$, and
the matrix $H(m,[u],\l)$) and generally speaking it has infinite
set of solutions. But if we fix a set of arguments of one of the
functions $H(m,[u],\l)$ or $F(m,[u])$ (i.e. if we fix number k or
M) we obtain additional conditions that suffice to determine the
desired functions. In the section 2 we represent several kinds of
boundary conditions compatible with zero curvature equation
(\ref{la}) of the discrete Toda chain (\ref{toda}). Some of them
was found earlier in \cite{su} and \cite{hk}.

The boundary condition (\ref{bc}) reduces the chain (\ref{toda})
to a half-line. To obtain finite-dimensional system we impose two
boundary conditions \bb\label{bcn}
u_{m,0}=F_1(m,u_{m,1},u_{m-1,1},...,u_{m,M},u_{m-1,M}),\quad
u_{m,N+1}=F_2(m,u_{m,1},u_{m-1,1},...,u_{m,K},u_{m-1,K}), \ee
$1\leq M,K\leq N$, which are assumed to be compatible with zero
curvature equation (\ref{la}). According our Definition above
equality (\ref{hl}) holds at the points $n_1=1$ and $n_2=N+1$,
while the functions $H(m,[u],\l)$ and $\tilde\l$ at these points
are equal to the matrices $H_1=H_1(m,[u],\l)$, $H_2=H_2(m,[u],\l)$
and involutions $\tilde\l_1$ and $\tilde\l_2$ respectively.

Let us suppose that involutions $\tilde\l_1$ and $\tilde\l_2$
coincide (i. e. $\tilde{\l}_1=\tilde{\l}_2=\tilde{\l}$). In this
case we can construct the generating function for the integrals of
motion of this system by the standard way \cite{sk}
\bb\label{pr_f}
   g(\lambda)=trace\left(P(m,\lambda)H^{-1}_{1}(m,\lambda)
   P^{-1}(m,\tilde\lambda)H_{2}(m,\lambda)\right),\ee
where $P(m,\lambda)= L_{m,N}(\lambda)\dots L_{m,1}(\lambda)$.
Similar to the continuous case \cite{sy} one can solve this system
by utilizing a definite  number of symmetries in addition to
integrals of motion (see \cite{kaz}). The set of symmetries needed
can be found by using the properly chosen master symmetries.

The case $\tilde{\l}_1\neq\tilde{\l}_2$ is considered in the
section 3. It is shown that if $N=1$ and
$\tilde{\l}_1\neq\tilde{\l}_2$ then the restricted chains for
certain choices closure conditions are reduced to the well known
discrete Painlev\'e equations $dP_{III}$, $dP_{V}$, $dP_{VI}$
(see (\ref{dp3}), (\ref{dp5}), (\ref{dp6}) below).

\section{Boundary conditions consistent with zero curvature equation.}
In this section we consider boundary condition of the form
(\ref{bc}) for the discrete Toda chain (\ref{toda}) assuming that
it's compatible with zero curvature equation. Let us start with
the matrix-equation (\ref{hl}) which is the main equation when
defining boundary conditions. It gives rise to a system of four
scalar equations on elements of matrices $H(m,\l)$ and $H(m+1,\l)$
(we denote $h_{ij}=(H(m,\l))_{ij}$ and
$\bar{h}_{ij}=(H(m+1,\l))_{ij}$)
\begin{eqnarray}
&&\l\bar{h}_{11}+\l\bar{h}_{12}F=\tilde{\l}h_{11}+h_{21}u_{m,1},\label{hl1}\\
&&\bar{h}_{11}u_{m,1}-\bar{h}_{12}=\tilde{\l}h_{12}+h_{22}u_{m,1},\label{hl2}\\
&&\l\bar{h}_{21}+\l\bar{h}_{22}F=\tilde{\l}h_{11}F-h_{21},\label{hl3}\\
&&\bar{h}_{21}u_{m,1}-\bar{h}_{22}=\tilde{\l}h_{12}F-h_{22}.\label{hl4}
\end{eqnarray}

{\bf Proposition 1.} {\it Suppose that the boundary condition
(\ref{bc}) for the discrete Toda chain (\ref{toda}) is compatible
with zero curvature equation (\ref{la}) and the corresponding
matrix $H=H(m,\l)$ depends on temporal variable $m$ and $\l$ only,
i. e. it doesn't depend on the dynamical variables. Then it reads
as \bb \label{bc1} \quad F={1\over u_{m,0}}=\alpha \mu
^{-2m}u_{m,1}+\beta\mu^{-m}. \ee }Here and below $\a$, $\b$, $\mu$
are arbitrary constants.

{\bf Remark.} We note that boundary condition (\ref{bc1}) was
previously found in the particular case when $\mu=1$ and $\a_1=0$,
$\b_1=0$, $\a_2=2$, $\b_2=0$ and $\a_3=0$, $\b_3=1$ (see
\cite{su}). Yu.B. Suris elaborated an algebraic structure of
finite-dimensional reductions of the discrete Toda chain
(\ref{toda}) obtained by imposing one of this boundary conditions.
In the case $u_{m,0}=\infty$, $u_{m,N+1}=-\infty$ complete
integrability of corresponding system are proved.

The case $\m=1$ and with arbitrary constants $\a$, $\b$ has been
investigated in \cite{kaz}. It was shown that the corresponding
finite-dimensional systems are integrated in quadratures.

{\bf Proof of Prop.1.} Since the elements of the matrix $H$ don't
depend on dynamical variables it follows from the equation
(\ref{hl2}) that $h_{12}=(-\tilde{\l})^ma$ where $a=const$ and
\bb\label{h111} \bar{h}_{11}=h_{22}. \ee If we assume that
$h_{12}\neq 0$ then the boundary condition $F$ is easily found
from (\ref{hl4})
$$
  F={\bar{h}_{21}u_{m,1}-\bar{h}_{22}+h_{22}\over\tilde{\l}h_{12}}.
$$
Substitution of expressions for $h_{12}$ and $F$ into (\ref{hl1})
yields
$$
 -\l\bar{h}_{21}u_{m,1}+\l\bar{h}_{22}=\tilde{\l}h_{11}+h_{21}u_{m,1}.
$$
In virtue of independence of the matrix $H$ on dynamical variables
the last equation gives $h_{21}=\left(-{1\over\l}\right)^mb$,
where $b=const$, and \bb\label{h221}
  \bar{h}_{22}={\tilde{\l}\over\l}h_{11}.
\ee Taking into account expressions (\ref{h111}), (\ref{h221}) and
independence of the function $F$ upon the parameter $\l$ we
immediately find the boundary condition (\ref{bc1}) where
notations $\a=-b$ and $\mu^2=1/a$ are used. The matrix $H$ and
involution $\tilde{\l}$ take the form \bb\label{h1}
  H(m,\lambda)=\left(\begin{array}{cc}
  (-1/\lambda)^{m-1}\frac{1}{\lambda+\mu}\beta\mu^{m-1} &
  (-1/\lambda)^{m}\mu^{2(m-1)}
  \\ -(-1/\lambda)^{m}\alpha &
  (-1/\lambda)^{m}\frac{1}{\lambda+\mu}\beta\mu^m
  \end {array}\right),\quad \tilde\lambda=\mu^2/\lambda.
\ee Proposition is proved.

{\bf Remark.}
 If $F=0$ , i. e. $\a=\b=0$, then we
have
$$
  H(m,\lambda)=\left(\begin{array}{cc} 0 & (-\lambda)^m\mu^{2(m-1)}  \\
  0 & 0\end {array}\right).
$$
In this case system (\ref{hl1})-(\ref{hl4}) has one more solution
$$
  H(m,\lambda)=\left(\begin{array}{cc} b & (-\lambda)^ma  \\
  0 & b\end {array}\right),\quad\tilde\lambda=\lambda.
$$

{\bf Proposition 2.}  {\it Suppose that the boundary condition
(\ref{bc}) for the discrete Toda chain (\ref{toda}) compatible
with zero curvature equation (\ref{la}) and the corresponding
matrix $H=H(m,u_{m,1},u_{m-1,1},\l)$ depends on dynamical
variables
 $u_{m,1}$ and $u_{m-1,1}$. Then it reads
as
\begin{eqnarray}
  && 1)\quad F={1\over u_{m,0}}=\mu^{-2m}{u_{m,1}u_{m,2}\over
  u_{m-1,1}}+{(\mu u_{m-1,1}-u_{m,1})^2\over
  u_{m-1,1}(\mu^{2m}-\mu^2u_{m,1}u_{m-1,1})}+\nonumber\\
  &&+{\a u_{m,1}+\b(\mu^{1-m}u^2_{m,1}+\mu^m)\over
  \mu^{2m}-\mu^2u_{m,1}u_{m-1,1}},\label{bc2}\\
  && 2) \quad F={1\over u_{m,0}}={u_{m,1}+u_{m,2}\over \a u^2_{m-1,1}}-{1\over
  u_{m,1}}.\label{bc3}
\end{eqnarray}}

{\bf Remark.} Consider the discrete Toda chain (\ref{toda}) with
boundary condition of the form (\ref{bc1}) where $\a=\b=0$ at the
left endpoint and with (\ref{bc2}) where $\m=1$, $\a=\b=0$ at the
right endpoint
\begin{eqnarray}
&& e^{-q_{m,0}}=0,\label{ld1}\\&&
q_{m+1,n}-2q_{m,n}+q_{m-1,n}=\ln\frac{e^{q_{m,n+1}-q_{m,n}}+1}{e^{q_{m,n}-q_{m,n-1}}+1},\quad
n=1,...,N-1, \label{ld2}\\&&
q_{m+1,N}-2q_{m,N}+q_{m-1,N}=\ln\frac{e^{q_{m-1,N}-2q_{m,N}-q_{m,N-1}}+
{(e^{q_{m-1,N}}-e^{q_{m,N}})^2\over
e^{2q_{m,N}}(e^{q_{m-1,N}+q_{m,N}}-1)
}+1}{e^{q_{m,n}-q_{m,n-1}}+1}.\label{ld3}\end{eqnarray} This
system in continuous limit corresponds to the generalized Toda
chain
\begin{eqnarray}&& e^{-q_{0}}=0,\label{d1}\\&&
q_{n,xx}=e^{q_{n+1}-q_{n}}-e^{q_{n}-q_{n-1}},\quad
n=1,..N-1,\label{d2}\\&& e^{q_{N+1}}=e^{-q_{N-1}}+{q^2_{N,x}\over
2shq_N},\label{d3}\end{eqnarray} which is related to the Lie
algebras of series $D_n$ \cite{ah}. Years ago in \cite{su} the
following discrete analogue of (\ref{d1})-(\ref{d3}) has been
suggested
\begin{eqnarray}
&& e^{-q_{m,0}}=0,\label{sld1}\\&&
q_{m+1,n}-2q_{m,n}+q_{m-1,n}=\ln\frac{e^{q_{m,n+1}-q_{m,n}}+1}{e^{q_{m,n}-q_{m,n-1}}+1},\quad
n=1,...,N-2, \label{sld2}\\&&
q_{m+1,N-1}-2q_{m,N-1}+q_{m-1,N-1}=\ln\frac{(e^{q_{m,N}-q_{m,N-1}}+1)(e^{-q_{m,N}-q_{m,N-1}}+1)}{e^{q_{m,N-1}-q_{m,N-2}}+1},
\label{sld3}\\&&
q_{m+1,N}-2q_{m,N}+q_{m-1,N}=\ln\frac{e^{-q_{m,N}-q_{m,N-1}}+1}{e^{q_{m,N}-q_{m,N-1}}+1}.\label{sld4}\end{eqnarray}
 Unfortunately we failed to find relation between these two discrete analogues.

{\bf Proof of Prop 2.} Consider the system of equations
(\ref{hl1})-(\ref{hl4}). Assume that $h_{12}\neq 0$ then it
follows from (\ref {hl4}) that
\begin{equation}\label{f}
  F ={\bar{h}_{21}u_{m,1}-\bar{h}_{22}+h_{22}\over\tilde{\l}h_{12}}.
\end{equation}

Let us differentiate equation (\ref {hl2}) with respect to the
variable $u_{m,2}$. This leads to the equation \bb\label{p22}
  \frac{\partial (\bar{h}_{11}u_{m,1}-\bar{h}_{12})}{\partial
  u_{m+1,1}}\frac{\partial u_{m+1,1}}{\partial u_{m,2}}=0.
\ee By setting $\frac{\partial u_{m+1,1}}{\partial u_{m,2}}\neq 0
$ (we can't find $F$ in the opposite case) and integrating
(\ref{p22}) one finds that
$\bar{h}_{11}u_{m,1}-\bar{h}_{12}=g_1(u_{m,1})$ or
$h_{11}=\frac{g_1(u_{m-1,1})+h_{12}}{u_{m-1,1}}$. Analysis of the
left-hand side of the equation (\ref {hl2}) leads us to expression
$h_{22}=\frac{g_1(u_{m,1})-\tilde{\l}h_{12}}{u_{m,1}}$. Here and
below we use
 $g_i(u_{m,1})$ to denote some functions depending on dynamical variables.

Substitute obtained expressions into (\ref {hl1})
\begin{eqnarray}
  && \l\bar{h}_{12}\bar{h}_{21}u_{m,1}-\l\bar{h}_{12}\frac{g_1(u_{m+1,1})}{u_{m+1,1}}+\l\tilde{\l}\bar{h}^2_{12}
  {1\over u_{m+1,1}}+\l\bar{h}_{12} {g_1(u_{m,1})\over
  u_{m,1}}=\nonumber\\
  && =\tilde{\l}^2 h_{12}\frac{g_1(u_{m-1,1})}{u_{m-1,1}}+\tilde{\l}^2
  h_{12}^2\frac{1}{u_{m-1,1}}+\tilde{\l}h_{12}h_{21}u_{m,1}-\l\tilde{\l}
  h_{12}\frac{g_1(u_{m,1})}{u_{m,1}}.\nonumber
\end{eqnarray}
Remind that $\frac{\partial u_{m+1,1}}{\partial u_{m,2}}\neq 0$,
so we can separate the last expression on two equalities
\begin{eqnarray}&&
  \l\bar{h}_{12}\bar{h}_{21}u_{m,1}-\l\bar{h}_{12}\frac{g_1(u_{m+1,1})}{u_{m+1,1}}+\l\tilde{\l}\bar{h}^2_{12}
  {1\over u_{m+1,1}}+\l\bar{h}_{12} {g_1(u_{m,1})\over
  u_{m,1}}=g_2(u_{m,1}),\label{m11}\\
&& \tilde{\l}^2
h_{12}\frac{g_1(u_{m-1,1})}{u_{m-1,1}}+\tilde{\l}^2
h_{12}^2\frac{1}{u_{m-1,1}}+\tilde{\l}h_{12}h_{21}u_{m,1}-\l\tilde{\l}
h_{12}\frac{g_1(u_{m,1})}{u_{m,1}}=g_2(u_{m,1}).\label{pr1}
\end{eqnarray}
Let us find $h_{21}$ from (\ref{pr1})
$$
h_{21}={g_2(u_{m,1})\over
\tilde{\l}h_{12}u_{m,1}}-{\tilde{\l}g_1(u_{m-1,1})\over
u_{m,1}u_{m-1,1}}-{\tilde{\l}h_{12}\over u_{m,1}u_{m-1,1}}+{\l
g_1(u_{m,1})\over u_{m,1}^2}.
$$
After that the equation (\ref {m11}) takes the form
\begin{eqnarray}
  && \bar{h}_{12}\left(\l{g_1(u_{m,1})\over
  u_{m,1}}-\l\tilde{\l}{g_1(u_{m,1})\over
  u_{m+1,1}}+\l^2{u_{m,1}g_1(u_{m+1,1})\over
  u_{m+1,1}^2}-\l{g_1(u_{m+1,1})\over u_{m+1,1}}\right)=\nonumber\\
  && =g_2(u_{m,1})-{\l \over \tilde{\l}}{g_2(u_{m+1,1})u_{m,1}\over
  u_{m+1,1}}\label{m12}.
\end{eqnarray}
One can easily see that the equation (\ref {hl3}) is rewritten by
means of (\ref {m12}) as follows
$$
\left({g_1(u_{m+1,1})\over u_{m+1,1}\bar{h}_{12}}-{\tilde{\l}\over
\l}{g_1(u_{m-1,1})\over
\bar{h}_{12}u_{m-1,1}}+\tilde{\l}{h_{12}\over
u_{m,1}\bar{h}_{12}}-\tilde{\l}{{1\over
u_{m+1,1}}}-{\tilde{\l}\over \l}{h_{12}\over
\bar{h}_{12}u_{m-1,1}}+{1\over u_{m,1}}\right)=0.
$$
The condition $\frac{\partial u_{m+1,1}}{\partial u_{m,2}}\neq 0$
allows one to obtain the following two equalities from the last
equation
\begin{eqnarray}
  && {g_1(u_{m+1,1})\over u_{m+1,1}}+\bar{h}_{12}{u_{m+1,1}-\tilde{\l}u_{m,1}\over u_{m+1,1}u_{m,1}}=g_3(u_{m,1}),\label{m31}\\
  && {\tilde{\l}\over \l}{g_1(u_{m-1,1})\over
  u_{m-1,1}}+h_{12}{\tilde{\l}\over \l}{u_{m,1}-\l u_{m-1,1}\over
  u_{m,1}u_{m-1,1}}=g_3(u_{m,1})\label{m32}.
\end{eqnarray}
We can find unknown $h_{12}$ from (\ref {m31})
\bb\label{h12}
   h_{12}={u_{m,1}u_{m-1,1}\over
   u_{m,1}-\tilde{\l}u_{m-1,1}}\left(g_3(u_{m-1,1})-{g_1(u_{m,1})\over
   u_{m,1}}\right).
\ee Substitution of this expression for $h_{12}$ into (\ref {m32})
leads to equality
\begin{eqnarray}
  &&{\l\over\tilde{\l}}g_3(u_{m,1})u_{m,1}+g_1(u_{m,1})+\tilde{\l}g_1(u_{m-1,1})+\l
  g_3(u_{m-1,1})u_{m-1,1}=\nonumber\\ &&=\l
  u_{m-1,1}g_3(u_{m,1})+u_{m,1}{g_1(u_{m-1,1})\over
  u_{m-1,1}}+g_3(u_{m-1,1})u_{m,1}+\l u_{m-1,1}{g_1(u_{m,1})\over
  u_{m,1} }.\label{m34}
\end{eqnarray}
Differentiation of (\ref {m34}) with respect to the variables
$u_{m,1}$ and $u_{m-1,1}$ gives equation
$$
{\partial\left({g_1(u_{m-1,1})\over
u_{m-1,1}}+g_3(u_{m-1,1})\right)\over\partial
u_{m-1,1}}=-\l{\partial\left({g_1(u_{m,1})\over
u_{m,1}}+g_3(u_{m,1})\right)\over\partial u_{m,1}},
$$
from which it follows that
$$
   g_3(u_{m,1})=\left(-{1\over
   \l}\right)^{m+1}c_0u_{m,1}-{g_1(u_{m,1})\over u_{m,1}}+c_1(m).
$$
Let $c_i=c_i(\l)$ and $c_i(m)=c_i(m,\l)$ be some functions
depending only on $\l$ and $\l$, $m$ respectively.

Substitute the expression for $g_3(u_{m,1})$ into (\ref {m34})
\begin{eqnarray}
  && {1\over\tilde{\l}}\left(-{1\over \l}\right)^m
  c_0u^2_{m,1}+{\l\over\tilde{\l}}g_1(u_{m,1})-{\l\over\tilde{\l}}c_1(m)u_{m,1}-g_1(u_{m,1})+c_1(m-1)u_{m,1}
  =\nonumber\\&& =-\left(-{1\over \l}\right)^{m-1} c_0u^2_{m-1,1}-\l
  g_1(u_{m-1,1})+\l c_1(m-1)u_{m-1,1}-\tilde{\l}g_1(u_{m-1,1})+\l
  c_1(m)u_{m-1,1}.\nonumber
\end{eqnarray}
Left and right-hand sides of the last equality depends only on
$u_{m,1}$ and $u_{m-1,1}$ respectively. Consequently
$c_1(m+1)={\tilde{\l}\over\l}c_1(m-1)$ and \bb\label{g1}
  g_1(u_{m,1})={1\over
  \tilde{\l}-\l}\left((-\tilde{\l})^{m+1}c_2+\left(-{1\over
  \l}\right)^m c_0 u^2_{m,1}-\l(c_1(m)-c_1(m+1))u_{m,1} \right).
\ee

Return to the equality (\ref {m12}). Taking into consideration
(\ref {h12}) and (\ref {g1}) one gets
\begin{eqnarray}
  && {g_2(u_{m,1})\over
  u_{m,1}}+\left(-{1\over\l}\right)^mc_0g_1(u_{m,1})+\l{g^2_1(u_{m,1})\over
  u^2_{m,1}}-\l c_1(m){g_1(u_{m,1})\over u_{m,1}}=\nonumber\\&&
  ={\l\over\tilde{\l}}{g_2(u_{m+1,1})\over
  u_{m+1,1}}+\left(-{1\over\l}\right)^{m+1}{\l\over\tilde{\l}}c_0g_1(u_{m+1,1})+
  {\l^2\over\tilde{\l}}{g^2_1(u_{m+1,1})\over
  u^2_{m+1,1}}-{\l^2\over\tilde{\l}} c_1(m+1){g_1(u_{m+1,1})\over
  u_{m+1,1}}.\nonumber
\end{eqnarray}
Analysis of the last equation shows that
$$
  g_2(u_{m,1})=\left({\tilde{\l}\over\l}\right)^mc_3u_{m,1}-\left(-{1\over\l}\right)^m
  c_0g_1(u_{m,1})u_{m,1}-\l{g^2_1(u_{m,1})\over u_{m,1}}+\l c_1(m)g_1(u_{m,1}).
$$

As the function $F$ doesn't depend on parameter $\l$ we have $
\tilde{\l}={\mu^2\over \l}$ and \bb\label{f}  F={g_2(u_{m,1})\over
  \mu^2\bar{h}_{12}h_{12}}-{1\over u_{m,1}},
\ee where
\begin{eqnarray}
  && h_{12}={\sqrt{\l}\over
  (\l^2-\mu^2)}\left(-{1\over\l}\right)^m(a_0u_{m,1}u_{m-1,1}+\mu^{2m}a_2),\nonumber\\
  && g_2(u_{m,1})={1\over (\l^2-\mu^2)^2}\left({1\over \l
  }\right)^{2m}(\mu^{2m}a_3u_{m,1}-\mu^2a^2_0u^3_{m,1}+\mu^{m+1}a_0a_1u^2_{m,1}-\nonumber\\&&-\mu^{4m+4}a^2_2{1\over
  u_{m,1}}-\mu^{3m+2}a_2a_1),\nonumber\\&&g_1(u_{m,1})={\sqrt{\l}\over(\l^2-\mu^2)}
  \left(-{1\over\l}\right)^m\left({\mu^{2m+2}a_2\over\l}-a_0u^2_{m,1}+{\mu^ma_1\over
  \l-\mu}u_{m,1}\right),\nonumber
\end{eqnarray} and $a_i$ are arbitrary constants.

The function $\bar{h}_{12}$ being contained in the expression for
$F$ depends on variable $u_{m+1,1}$ which is not dynamical, i. e.
it can be expressed through variables $u_{m,1}$, $u_{m-1,1}$ and
function $F$
$$
  u_{m+1,1}={(u_{m,1}+u_{m,1})u_{m,1}\over u_{m-1,1}(1+u_{m,1}F)}.
$$
Therefore if we denote $a_0=-a_2\mu^2$,
$a_3=a_2\mu^4(2a_2\mu+\a)$, $a_1=\b a_2\mu^2$ then the equality
(\ref {f}) gives boundary condition (\ref{bc2}). The matrix $H$
and involution $\tilde{\l}$ are following
$$
  H=\left(\begin{array}{cc}
  {g_1(u_{m-1,1})+h_{12}\over u_{m-1,1}}&
  h_{12}\\
  {\l g_2(u_{m,1})\over
  \mu^2h_{12}u_{m,1}}-{\mu^2(g_1(u_{m-1,1})+h_{12})\over \l
  u_{m,1}u_{m-1,1}}+\l{g_1(u_{m,1})\over u^2_{m,1}}& {\l
  g_1(u_{m,1})-\mu^2h_{12}\over\l u_{m,1}}
  \end {array}\right),\quad \tilde{\l}={\mu^2\over \l}.
$$

Now suppose that $h_{12}=0$. Then the system
(\ref{hl1})-(\ref{hl4}) takes the form
\begin{eqnarray}
  &&\l\bar{h}_{11}=\tilde{\l}h_{11}+h_{21}u_{m,1},\label{mhl1}\\
  &&\bar{h}_{11}=h_{22},\label{mhl2}\\
  &&\l\bar{h}_{21}+\l\bar{h}_{22}F=\tilde{\l}h_{11}F-h_{21},\label{mhl3}\\
  &&\bar{h}_{21}u_{m,1}-\bar{h}_{22}=h_{22}.\label{mhl4}
\end{eqnarray}
The system (\ref{mhl1})-(\ref{mhl4}) has a nontrivial solution if
we assume that ${\partial u_{m+1,1}\over\partial u_{m,2}}=0$, i.
e.
$$
  F=g_0(u_{m,1},u_{m-1,1})\left(1+{u_{m,2}\over u_{m,1}}\right)-{1\over u_{m,1}},
$$
and consequently
$$
  u_{m+1,1}={u_{m,1}\over u_{m-1,1}g_0(u_{m,1},u_{m-1,1})}.
$$

Taking into account (\ref{mhl2}) we get from the equations
(\ref{mhl1}) and (\ref{mhl4})
$$
  \bar{h}_{11}=h_{11}{\tilde{\l}u_{m-1,1}-u_{m,1}\over \l
  u_{m-1,1}-u_{m,1}},\quad h_{21}=h_{11}{\tilde{\l}-\l\over \l
  u_{m-1,1}-u_{m,1}},
$$
and so (\ref{mhl3}) takes the form \bb\label{eq}
  (\l-\tilde{\l})(1+u_{m,1}F)(u_{m+1,1}-\l\tilde{\l}u_{m-1,1})=0.
\ee It implies that $u_{m+1,1}=\l\tilde{\l}u_{m-1,1}$. The other
factors in (\ref{eq}) don't vanish. Really, if $\l-\tilde\l=0$
then $H$ is equal to the identity matrix, which gives no
involution. As for the middle factor it coincides with the factor
$1+{u_{m,1}\over u_{m,0}}$ which is contained in the denominator
of the chain itself. In the domain of the right hand side  of the
chain (\ref{toda}) it must be different from zero.
 Since ${\partial
u_{m+1,1}\over\partial \l}=0$ we have $\tilde{\l}={\a\over\l}$.
Thus, boundary condition $F$ takes the form (\ref{bc3}), the
matrix $H$ and the involution $\tilde{\l}$ respectively are of the
form
$$
  H=\left(\begin{array}{cc} g(m)& 0 \\
  g(m){\a-\l^2\over\l(\l u_{m-1,1}-u_{m,1})} & g(m+1)
   \end {array}\right),\quad \tilde{\l}={\a\over \l},
$$
where $g(m)=\prod\limits^m_{k=0}{\a u_{m-1,1}-\l u_{m,1}\over\l(\l
u_{m-1,1}-u_{m,1})}$. The proposition is proved.

\section{Discrete Painlev\'e equations}

Consider the truncated system (\ref{toda}), (\ref{bcn}) in the
case of different involutions $\tilde{\l}_1\neq \tilde{\l}_2$ at
the endpoints $n=0$ and $n=2$. So that the endpoints are taken as
close as possible, i. e. $N=1$. The boundary conditions $F_1$ and
$F_2$ imposed at $n=0$ and $n=2$ are of one of the forms
represented  by (\ref{bc1}) or (\ref{bc2}). Denote through
$H_1(\l,m)$, $H_2(\l,m)$ the matrices $H$ corresponding to each
endpoint. In the case of (\ref{bc1}) and (\ref{bc2}) the
involutions are of the form $\tilde{\l}_1={\mu^2_1\over \l}$,
$\tilde{\l}_2={\mu^2_2\over \l}$. Thus the system (\ref{toda}),
(\ref{bcn}) takes the form
\begin{eqnarray}
&&{1\over u_{m,0}}=F_1(m,u_{m,1},u_{m-1,1}),\label{dpf1}\\
&&u_{m+1,1}={u^2_{m,1}(1+u_{m,2}/u_{m,1})\over
u_{m-1,1}(1+u_{m,1}/u_{m,0})},\label{dtoda}\\
&& u_{m,2}=F_2(m,u_{m,1},u_{m-1,1}).\label{dpf2}
\end{eqnarray}
It was shown in \cite{ah} that the differential-difference Toda
equation (\ref{ut}) admits finite dimensional reductions of the
Painlev\'e type. The same can be done in our case of purely
discrete equations.

{\bf Proposition 3.} {\it The system (\ref{dpf1})-(\ref{dpf2}) is
equivalent to the matrix equation \bb\label{dpma}
  A_m\left(\d\lambda\right)M_m(\lambda)=M_{m+1}(\lambda)A_m(\lambda),
\ee which is the consistency condition of two linear equations
\begin{eqnarray}
  && Y_{m+1}(\lambda)=A_m(\lambda)Y_m(\lambda),\label{pa}\\
  && Y_{m}\left(\d\lambda\right)=M_m(\lambda)Y_m(\lambda),\label{pm}
\end{eqnarray}
where
$M_m(\lambda)=H_{1}\left(\frac{\mu^2_2}{\lambda},m\right)L^{-1}_{m}\left(\frac{\mu^2_2}{\lambda}\right)H^{-1}_{2}
\left(\frac{\mu^2_2}{\lambda},m\right)L_{m}(\lambda)$  and
$\d=\frac{\mu^2_1}{\mu^2_2}$.}

{\bf Proof.} Boundary conditions (\ref{dpf1}) and (\ref{dpf2}) are
consistent with zero curvature equation (\ref{la}). It follows
from it that equation (\ref{a}) taken at the spatial points $n=1$
and $n=2$
\begin{eqnarray}
  && Y_{m+1,1}(\lambda)=A_{m,1}(\lambda)Y_{m,1}(\lambda), \label{a0}\\
  && Y_{m+1,2}(\lambda)=A_{m,2}(\lambda)Y_{m,2}(\lambda)\nonumber
\end{eqnarray}
possesses additional linear transformations
\begin{eqnarray}
  && Y_{m,1}\left(\frac{\mu^2_1}{\lambda}\right)=H_{1}(\lambda,m)Y_{m,1}(\lambda),\label{sy1}\\
  && Y_{m,2}\left(\frac{\mu^2_2}{\lambda}\right)=H_{2}(\lambda,m)Y_{m,2}(\lambda).\label{sy2}
\end{eqnarray}
As we said above the system (\ref{dpf1})-(\ref{dpf2}) is
equivalent to the consistency condition of the equation (\ref{a0})
with the following one\bb\label{l0}
  Y_{m,2}(\lambda)=L_{m,1}(\lambda)Y_{m,1}(\lambda).
\ee Replacing $\l\rightarrow {\mu^2_2\over\lambda}$ in (\ref{l0})
and taking to account (\ref{sy2}) gives
$$
  Y_{m,1}\left(\frac{\mu^2_2}{\lambda}\right)=L^{-1}_{m,1}\left(\frac{\mu^2_2}{\lambda}\right)H_{2}(\lambda,m)
  L_{m,1}(\lambda)Y_{m,1}(\lambda).
$$
Substitute the last expression into (\ref{sy1}) \bb\label{hy}
  Y_{m,1}\left(\frac{\mu^2_1}{\lambda}\right)=H_{1}(\lambda,m)L^{-1}_{m,1}(\lambda)H^{-1}_{2}(\lambda,m)
  L_{m,1}\left(\frac{\mu^2_2}{\lambda}\right)Y_{m,1}\left(\frac{\mu^2_2}{\lambda}\right).
\ee Replacing again $\lambda \rightarrow \frac{\mu^2_2}{\lambda}$
in (\ref{hy})  we get the equality
$$
  Y_{m,1}\left(\frac{\mu^2_1}{\mu^2_2}\lambda\right)=H_{1}\left(\frac{\mu^2_2}{\lambda},m\right)L^{-1}_{m,1}
  \left(\frac{\mu^2_2}{\lambda}\right)H^{-1}_{2}\left(\frac{\mu^2_2}{\lambda},m\right)
  L_{m,1}(\lambda)Y_{m,1}(\lambda).
$$
Omit the second subindex in $u_{m,1}$. So the equation (\ref{l0})
is equivalent to the equation (\ref{pm}). The proposition is
proved.

Thus the system (\ref{dpf1})-(\ref{dpf2}) possesses Lax pair
(\ref{pa}), (\ref{pm}), which is typical for the discrete
Painlev\'e equations. Consider several illustrative examples.

{\bf Example 1.} The system (\ref{dpf1})-(\ref{dpf2}) with
boundary conditions
$$
  {1\over u_{m,0}}=\a_1u_{m,1}+\b_1,\quad
  u_{m,2}=\a_2\mu^{2m}{1\over u_{m,1}}+\b_2\mu^m
$$
gives rise the equation on variables $u_m=u_{m,1}$ \bb\label{dp3}
  u_{m+1}u_{m-1}={u^2_m+\beta_2\mu^mu_m+\alpha_2\mu^{2m}\over\alpha_1
  u^2_m+\beta_1u_m+1},
\ee which is one of the forms of the third discrete Painlev\'e
equation $d-P_{III}$ \cite{rgh}, \cite{rg1}. Check that in the
continuous limit it approaches the $P_{III}$ equation. Return to
the variables $u_{m}=e^{q_{m}}$ and take $\mu=e^{2h}$,
$\a_1=\bar{\a_1}h^2$, $\a_2=\bar{\a_2}h^2$, $\b_1=\bar{\b_1}h^2$,
$\b_2=\bar{\b_2}h^2$. Then the equation (\ref{dp3}) takes the form
$$ q_{m+1}-2q_m+q_{m-1}=ln{1+h^2(\bar\a_2e^{4mh-2q_m}+\bar{\b}_2e^{2mh-q_m})\over 1+h^2(\bar{\a}_1e^{2q_m}+\bar\b_1e^{q_m})}.$$
Let $h\rightarrow 0$ in the last equation then we have
\bb\label{p3}
q_{xx}=\bar{\a}_2e^{4x-2q}+\bar{\b}_2e^{2x-q}-\bar{\a}_1e^{2q}-\bar{\b}_1e^q.\ee
Substitution $e^{q(x)}=zy(z)$, $z=e^x$ in (\ref{p3}) gives the
third Painlev\'e equation \cite{p}
$$
y_{zz}=\frac{y^2_z}{y}-\frac{y_z}{z}+\frac{1}{z}(Ay^2+B)+Cy^3+\frac{D}{y},
$$
where parameters are $A=-\bar{\b}_1$, $B=\bar{\b}_2$,
$C=-\bar{\a}_1$, $D=\bar{\a}_2$.

By using Prop.3 we can find a matrix $M$ for zero curvature
equation (\ref{dpma}) according to the equation (\ref{dp3}) (we
denote $m_{ij}=(M)_{ij}$)
\begin{eqnarray}
&& m_{12}={1\over \varphi}(\mu^{m+1}\l\b_2-\a_2(\mu+\l)\xi
u_{m}),\nonumber\\
&& m_{11}=m_{12}\left({\l\over u_{m}}+{1\over
u_{m-1}}\right)+{\mu^m\l\over\varphi
u_{m-1}}\left(\b_2\xi-\mu^m(\mu+\lambda){1\over
u_m}\right),\nonumber\\
&& m_{22}={1\over \varphi}\left(\b_1\b_2\mu^{m+1}{\l^2\over
\mu^2+\l}-\a_2(\mu+\l)\eta u_{m}\right),\nonumber\\
&& m_{21}=m_{22}\left({\l\over u_m}+{1\over
u_{m-1}}\right)+{\mu^m\l\over\varphi
u_{m-1}}\left(\b_2\eta-\mu^m(\mu+\l)\b_1{\l\over
u_m(\mu^2+\l)}\right),\nonumber
\end{eqnarray}
where
$$
 \varphi={\mu^{2m}\over \mu+\l}(\a_2(\mu+\l)^2-\mu\l\b^2_2),
$$
$$
\xi={\mu^2\l\b_1u_{m-1}\over \mu^2+\l}+{\mu^2 u_{m-1}+\l u_m\over
u_m}, \quad \eta=\a_1\l u_{m-1}+{\b_1\l(\mu^2u_{m-1}+\l u_m)\over
u_m(\mu^2+\l)}.
$$

{\bf Example 2.} Impose boundary condition (\ref{bc1}) at the
point $n=1$
$$
   {1\over u_{m,0}}=\a_1\mu^{-2m}u_{m,1}+\b_1\mu^{-m},$$
and (\ref{bc2}) where $\mu=1$ at the point $n=2$ $$
u_{m,2}={u_{m-1,1}\over
  u_{m,1}u_{m,0}}-{(u_{m-1,1}-u_{m,1})^2\over
  u_{m,1}(1-u_{m,1}u_{m-1,1})}+{(\a_2(u^2_{m,1}+1)+\b_2u_{m,1})u_{m-1,1}\over
  u_{m,1}(1-u_{m,1}u_{m-1,1})}.$$
Under this constraints
  the Toda chain (\ref{toda}) is reduced to
the fifth discrete Painlev\'e equation $d-P_V$ \cite{rg2}
\bb\label{dp5}
  (u_{m+1}u_m-1)(u_mu_{m-1}-1)={pq(u_m-a)(u_m-1/a)(u_m-b)(u_m-1/b)\over(u_m-p)(u_m-q)},
\ee where $p=p_0\mu^m$, $q=q_0\mu^m$ and $p_0$, $q_0$, $a$, $b$
are constants such as following equalities have place
\begin{eqnarray}
  && p_0q_0=\a_2,\quad p_0+q_0=-\b_2,\nonumber\\
  && a+{1\over a}+b+{1\over b}=\a_1,\quad \left(a+{1\over
  a}\right)\left(b+{1\over
  b}\right)=-(3+\b_1).\nonumber
\end{eqnarray}

Return to the variables $u_m=e^{q_m}$ in (\ref{dp5}) again and
take $\mu=e^{-h}$. We use the same constants $\a_1$, $\a_2$,
$\b_1$ and $\b_2$ as in example 1. If $h\rightarrow 0$ then we
have an equation
$$
q_{xx}=\bar\a_1e^{2x}(1-e^{2q})+\bar\b_1e^x(e^{-q}-e^q)+{q^2_x\over
e^{2q}-1}+{\bar\a_2(e^q+e^{-q})+\bar\b_2\over 1-e^{2q}},
$$
which gives the fifth Painlev\'e equation by substitution
$e^{q(x)}={y(z)+1\over y(z)-1}$, $z=e^x$ \cite{p}
$$
y_{zz}=\left({1\over 2y}+{1\over y-1}\right)y^2_z-{y_z\over
z}+{(y-1)^2\over z^2}\left(Ay+{B\over y}\right)+C{y\over
z}+D{y(y+1)\over y-1}.
$$
Here parameters are follow $8A=-\bar\b_2-2\bar\a_2$,
$8B=\bar\b_2-2\bar\a_2$, $C=-2\bar\b_1$, $D=-2\bar\a_1$.

We will use following notation
\begin{eqnarray}
 && h(\l,\mu)={\sqrt{\l}\over
 \l^2-\mu^2}(\mu^{2m}-\mu^2u_mu_{m-1}),\nonumber\\
 && g(u_m,\l,\mu,\beta)={\sqrt{\l}\over
 \l^2-\mu^2}\left({\mu^{2m+2}\over\l}+\mu^2u^2_m+\b\mu^{m+2}{u_m\over\l-\mu}\right),\nonumber\\
 && f(u_m,\l,\mu,\a,\b)={1\over(\l^2-\mu^2)^2}\left(\mu^{2m+4}u_m(2\mu+\a)-\mu^6u^3_m-\mu^{m+5}\b u^2_m-
 {\mu^{4m+4}\over u_m}-\mu^{3m+4}\b\right).\nonumber
\end{eqnarray}
In this example functions $h$, $g(u_m)$ and $f(u_m)$ correspond to
following functions $h({1\over\l},1)$, $g(u_m,{1\over\l},1,\b_1)$
and $f(u_m,{1\over\l},1,\a_1,\b_1)$. Therefore elements of the
matrix $M$ take the form
\begin{eqnarray}
&&
m_{12}={u_m\over\varphi}(\mu^{2m-2}\l\xi_2 u_{m-1}+\eta\zeta),\nonumber\\
&& m_{11}={m_{12}}\left({\l\over u_m}+{1\over
u_{m-1}}\right)-{\l\over\varphi u_{m-1}}(\mu^{2m-2}\l hu_mu_{m-1}+\eta\xi_1\l u^2_m),\nonumber\\
&& m_{22}={u_m\over \varphi}\left({\l^2\mu^m\b_1\xi_2 u_{m-1}\over 1+\l\mu}+{\zeta\psi\over h}\right),\nonumber\\
&& m_{21}=m_{22}\left({\l\over u_m}+{1\over
u_{m-1}}\right)-{\l\over\varphi u_{m-1}}(\l\mu^m\b_1
hu_mu_{m-1}+\psi\xi_1\l u^2_m),\nonumber
\end{eqnarray}
where
\begin{eqnarray}
&& \xi_1=g(u_{m-1})+h,\quad \xi_2=g(u_m)-\l h,\nonumber\\
&&\varphi=\l\xi_1u_mg(u_m)-u_{m-1}(f(u_m)u_m+g(u_m)h),\quad
\eta={\l\b_1\mu^{m-1}u_{m-1}\over 1+\l\mu}+{\mu^{2m-2}(u_{m-1}+\l
u_m)\over u_m},\nonumber\\
&&\zeta=f(u_m)u_mu_{m-1}-\l^2h\xi_1u_m+g(u_m)hu_{m-1},\quad
\psi=\a_1\l u_{m-1}+{\l\mu^m\b_1(u_{m-1}+\l u _m)\over
u_m(1+\l\mu)}.\nonumber
\end{eqnarray}

{\bf Example 3.} Consider the chain (\ref{toda}) with boundary
conditions (\ref{bc2}) where $\mu$ is arbitrary constant at the
point $n=0$
$$
  {1\over u_{m,0}}=\mu^{-2m}{u_{m,1}u_{m,2}\over
  u_{m-1,1}}+{(\mu u_{m-1,1}-u_{m,1})^2\over
  u_{m-1,1}(\mu^{2m}-\mu^2u_{m,1}u_{m-1,1})}+{\a_1(\mu^{1-m}u^2_{m,1}+\mu^m)+\b_1u_{m,1}\over
  \mu^{2m}-\mu^2u_{m,1}u_{m-1,1}},$$
  and where $\mu=1$ at the $n=2$
  $$ u_{m,2}={u_{m-1,1}\over
  u_{m,1}u_{m,0}}-{(u_{m-1,1}-u_{m,1})^2\over
  u_{m,1}(1-u_{m,1}u_{m-1,1})}+{(\a_2(u^2_{m,1}+1)+\b_2u_{m,1})u_{m-1,1}\over
  u_{m,1}(1-u_{m,1}u_{m-1,1})}.$$
Solving these equations with regard to the variables $u_{m,2}$ and
$u_{m,0}$ and substituting them into (\ref{toda}) one gets
 an equation on variables $u_m=u_{m,1}$ which is the sixth discrete
Painlev\'e equation $d-P_{VI}$ \cite{rg3} \bb\label{dp6}
  {(u_{m+1}u_m-p_{m+1}p_m)(u_mu_{m-1}-p_mp_{m-1})\over (u_{m+1}u_m-1)(u_mu_{m-1}-1)}=
  {(u_m-ap_m)(u_m-p_m/a)(u_m-bp_m)(u_m-p_m/b)\over(u_m-c)(u_m-1/c)(u_m-d)(u_m-1/d)},
\ee where $p=p_0\mu^m$, $p^2_0=1/\mu$ and $a$, $b$, $c$, $d$ are
constants satisfying the following conditions
\begin{eqnarray}
  &&  a+{1\over a}+b+{1\over b}=-{\a_1\over\mu p_0},\quad
  \left(a+{1\over a}\right)\left(b+{1\over
  b}\right)=4+{\b_1\over\mu},\nonumber\\ && c+{1\over c}+d+{1\over
  d}=\a_2,\quad \left(c+{1\over c}\right)\left(d+{1\over
  d}\right)=-(4+\b_2).\nonumber
\end{eqnarray}
One can take $\mu=e^h$ to get the continuous limit
\begin{eqnarray}&& q_{xx}={e^{-q}-e^q\over
(1-e^{2x})(e^{q-2x}-e^{-q})}((q_x-1)^2+\bar\a_1(e^{q-x}+e^{x-q})+\bar\b_1)-\nonumber\\&&
-{e^{2x-q}-e^q\over
(1-e^{2x})(e^q-e^{-q})}(q^2_x-\bar\a_2(e^q+e^{-q})-\bar\b_2).\nonumber\end{eqnarray}
Substitution $e^{q(x)}={y(z)+\sqrt{z}\over y(z)-\sqrt{z}}$,
$e^x={1+\sqrt{z}\over 1-\sqrt{z}}$ gives at once the sixth
Painlev\'e equation \cite{p}
\begin{eqnarray}
&& y_{zz}={1\over 2}\left({1\over y}+{1\over y-1}+{1\over
y-z}\right)y^2_z-\left({1\over z}+{1\over z-1}+{1\over
y-z}\right)y_z+\nonumber\\&& +{y(y-1)(y-z)\over
z^2(z-1)^2}\left(A+B{z\over y^2}+C{z-1\over
(y-1)^2}+D{z(z-1)\over(y-z)^2}\right),\nonumber
\end{eqnarray}
where parameters are following $8A=\bar\b_2+2\bar\a_2$,
$8B=-\bar\b_2+2\bar\a_2$, $8C=-\bar\b_1-2\bar\a_1$,
$8D=\bar\b_1-2\bar\a_1+4$.

Elements of the matrix $M$ for the equation (\ref{dp6}) have the
form
\begin{eqnarray}
  && m_{12}={u_m\over\varphi}(\l
  h^2_1\psi_2u_{m-1}-\xi_1\zeta),\nonumber\\
  && m_{11}=m_{12}\left({\l\over u_m}+{1\over u_{m-1}}\right)+{\l^2h_1u_m\over\varphi
  u_{m-1}}(\xi_1\psi_1-h_1h_2u_{m-1}),\nonumber\\
  && m_{22}={u_m\over\varphi}\left({\l h_1u_{m-1}\over
  u_m}\psi_2\xi_2-\eta\zeta\right),\nonumber\\
  && m_{21}=m_{22}\left({\l\over u_m}+{1\over u_{m-1}}\right)+{\l^2 h_1\over\varphi
  u_{m-1}}(\eta\psi_1 u_m-h_2\xi_2 u_{m-1}),\nonumber
\end{eqnarray}
where we denote
\begin{eqnarray}
  && h_1=h(1/\l,\mu),\quad h_2=h(1/\l,1),\quad
  g_1(u_m)=g(u_m,1/\l,\mu,\b_1),\quad
  g_2(u_m)=g(u_m,1/\l,1,\b_2),\nonumber\\
  && f_1(u_m)=f(u_m,1/\l,\mu,\a_1,\b_1),\quad
  f_2(u_m)=f(u_m,1/\l,1,\a_2,\b_2),\nonumber\\
  && \xi_1=\l u_mg_1(u_{m-1})-h_1u_{m-1},\quad
  \xi_2=g_1(u_m)-\mu^2\l h_1,\nonumber\\
  && \psi_1=g_2(u_{m-1})+h_2,\quad \psi_2=g_2(u_m)-\l
  h_2,\nonumber\\
  && \eta={f_1(u_m)u_{m-1}\over h_1\mu^2}-\l g_1(u_m)-{\l\mu^2\xi_1\over
  u_m},\quad \zeta=f_2(u_m)u_{m-1}-\l^2 h_1\psi_1+h_1g_2(u_m){u_{m-1}\over u_m},\nonumber\\
  && \varphi=\l h_1
  \psi_1g_2(u_m)u-h_2u_{m-1}(f_2(u_m)u_m+g_2(u_m)h_1).\nonumber
\end{eqnarray}

\paragraph{\large Acknowledgments.} The work was supported by RFBR
grant \# 01-01-00931.

\end{document}